\documentclass[11pt,fleqn]{article}

\usepackage{amsfonts,amssymb,amsmath}
\usepackage{epsf,epsfig,graphics,graphicx}
\usepackage[english,russian]{babel}
\usepackage{xcolor}

\usepackage{amsfonts,amssymb,cite}
\usepackage{epsf,graphicx}

\def\noi{\noindent}
\def\jnumber#1#2{\thispagestyle{empty} \noi\unitlength=1mm
    	\begin{picture}(178,10)
            \put(177,15){\llap{\large\it Grav. Cosmol. No.\,#1, #2}}
                    \end{picture}}

\newcommand{\Title}[1]{\noi {{\Large\bf #1}}\\[1ex]}

\def\Aunames#1{\noi{\bf #1}}
\def\auth#1{${}^{#1}$}
\def\Addresses#1{\medskip\noi \protect
	\begin{description}\itemsep -3pt {\it #1} \end{description}}
\def\addr#1#2{\item[${}^{#1}$]{\it #2}}

\newcommand{\Rec}[1]{\noi {\it Received #1} \\}

\newcommand{\Abstract}[1]{\vskip 2mm \begin{center}
        \parbox{16.4cm}{\small\noi #1} \end{center}\medskip}
\newcommand{\PACS}[1]{\begin{center}{\small PACS: #1}\end{center}}

\def\email#1#2{\footnotetext[#1]{e-mail: #2}\addtocounter{footnote}{1}}

\def\nqq{\hspace*{-2em}}

\def\Jl#1#2{#1 {\bf #2},\ }

\def\ApJ#1 {\Jl{Astroph. J.}{#1}}
\def\CQG#1 {\Jl{Class. Quantum Grav.}{#1}}
\def\DAN#1 {\Jl{Dokl. AN SSSR}{#1}}
\def\GC#1 {\Jl{Grav. Cosmol.}{#1}}
\def\GRG#1 {\Jl{Gen. Rel. Grav.}{#1}}
\def\JETF#1 {\Jl{Zh. Eksp. Teor. Fiz.}{#1}}
\def\JETP#1 {\Jl{Sov. Phys. JETP}{#1}}
\def\JHEP#1 {\Jl{JHEP}{#1}}
\def\JMP#1 {\Jl{J. Math. Phys.}{#1}}
\def\NPB#1 {\Jl{Nucl. Phys. B}{#1}}
\def\NP#1 {\Jl{Nucl. Phys.}{#1}}
\def\PLA#1 {\Jl{Phys. Lett. A}{#1}}
\def\PLB#1 {\Jl{Phys. Lett. B}{#1}}
\def\PRD#1 {\Jl{Phys. Rev. D}{#1}}
\def\PRL#1 {\Jl{Phys. Rev. Lett.}{#1}}

\def\lal{&&\nqq {}}

\def\beq{\begin{equation}}
\def\eeq{\end{equation}}
\def\bear{\begin{eqnarray}}
\def\bearr{\begin{eqnarray} \lal}
\def\ear{\end{eqnarray}}
\def\earn{\nonumber \end{eqnarray}}

\begin{document}
\twocolumn[
\jnumber{}{2017}

\vspace{-4mm}

\Title{Exact Inflation in Einstein-Gauss-Bonnet Gravity}

\Aunames {I.V. Fomin,\auth{a,1} and S.V. Chervon\auth{b,c,2} }

\Addresses{
\addr a {Bauman Moscow State Technical University,
  2-nd Baumanskaya street, 5, Moscow, 105005, Russia}
\addr b {Ulyanovsk State Pedagogical University, Ulyanovsk, 100-years V.I. Lenin's Birthday Square, B.4, 432071, Russia}
\addr c {Astrophysics and Cosmology Research Unit,
School of Mathematics, Statistics and Computer Science,
University of KwaZulu-Natal, Private Bag X54 001,
Durban 4000, South Africa}
}
\Rec {10 April, 2017}

\Abstract
{We study cosmological inflation in the Einstein gravity model, where additionally the Gauss-Bonnet term non-minimally coupled to a scalar field is included. We prove that inflationary solutions of exponential and power-law types are allowable and we found few examples of them.
We also proposed the method of the exact inflationary solutions construction for a single scalar field with given scale factor and Gauss-Bonnet coupling term in the spatially flat Friedmann-Robertson-Walker Universe on the basis of connection with standard inflation and using special anzatses. With one special anzats we presented the system of equations in the form which allowed generation of exact solutions (at least in quadratures) of wide class by setting the scale factor.

}

 ]
\email 1 {ingvor@inbox.ru}
\email 2 {chervon.sergey@gmail.com}

\begin{otherlanguage}{english}

\section{Introduction}

The theory of cosmological inflation successfully describes a condition of the accelerated expansion of the Universe evolution at early stages~\cite{Starobinsky}. Also inflationary cosmology explains an origin of primary inhomogeneities and predicts their spectrum ~\cite{Mukhanov, Liddle}.

According to the theory of inflation
primordial perturbations happened from quantum fluctuations.
These fluctuations had essential amplitudes in scales of planck length and during inflation they generate the primordial perturbations which then lead nearer to scales of galaxies with almost same amplitudes.
Thus, inflation connects large-scale structure of the Universe with microscopic scales. The resultant range of inhomogeneities
practically doesn't depend on private scenarios of inflation and has a universal form. It leads to unambiguous predictions for a range of anisotropy of the background radiation ~\cite{Liddle}.

Models of inflation are set by a shape
of an effective potential $V(\phi)$. In this case, a scalar field $\phi$ comes down to $V(\phi)$ minimum in the inflationary stage. The end of inflation leads to violation of conditions of slow rolling, the field oscillates about a minimum and process of post-inflationary heating begins. This process includes at once some various stages, such as disintegration of inflaton condensate, the birth of particles of standard model and their thermalization \cite{Mukhanov,Liddle}.

Cosmological acceleration specifies that now in the Universe evenly distributed slowly changing space liquid with negative pressure, called by dark energy, dominates~\cite{Perlmutter}.

To make the specification of various types of cosmic fluid the phenomenological ratio $p=w\varepsilon$ between pressure $p$ and energy density  $\varepsilon$ with the state parameter $w$ for each species of the fluid is usually used.
Modern experiments~ \cite{Planck} testify that the Universe is spatially flat and at present time the state parameter for dark energy is $w =-1\pm0.1$.

Standard way of obtaining time-dependent parameter of a state is inclusion of scalar fields in cosmological model. At rather general assumptions within four-dimensional spacetime with a single scalar field as the source of gravity it was obtained the restrictions on state parameter for quintessence $-1<w<-\frac{1}{3}$ and for phantom scalar field $w<-1$~\cite{Liddle}.

Despite the fact that the inflationary scenario solves the problem of the big bang theory, for instance, the horizon and flatness problem, there are still unsolved problems such as initial singularity problem and quantum gravity.

For the very early Universe approaching the Planck scale, we could consider Einstein gravity with some corrections as the effective theory of the quantum gravity. The effective supergravity action from superstrings induces correction terms of higher order in the curvature, which may play a significant role in the early Universe. The one of such correction is the Gauss-Bonnet (GB) term in the low-energy effective action of the heterotic string~\cite{Zwiebach, gos87}.

Also, the GB term arises in the second order of Lovelock gravity, which is the generalization of the Einstein gravity. The Lovelock  gravity is described by the action of a special kind, consisting of the sum of the so-called Lovelock tensors~\cite{Lovelock}. The first order theory of Lovelock coincides with Einstein gravity, the second order is Einstein-Gauss-Bonnet (EGB) gravity, which will contain GB term and the variation of this term is different from zero only in a space with dimensions of not less than five. However, in the four-dimensional space, in the case of non-minimal interaction of a scalar field with GB term, the equations of the dynamics are quite different from the standard inflation~\cite{Guo, Koh}.

The inclusion of a scalar field in the inflationary model implies, in particular, consideration of the Higgs mechanism in connection with the experimental discovery of the Higgs boson~\cite{ATLAS}. Inflation can be generated by the Higgs boson, if it is non-minimally connected with the scalar Ricci with (negative) large coupling constant~\cite{Bezrukov}. The existence of a conformal transformation in such models leads to the fact that the amplitude of cosmological perturbations can be increased or reduced by adjusting this constant to match the observational data. Standard inflation Higgs mechanism makes it possible to consider other types of interaction of a scalar field with gravity (with the GB term) in the context of the Higgs field as inflaton~\cite{Cai}.

Thus, cosmological inflation, in the case of Lovelock gravity in the second order with non-minimal coupling of a scalar field with Gauss-Bonnet term for the four-dimensional spatially flat Friedmann-Robertson-Walker (FRW) Universe associated with both: low-energy limit of string theory and a generalization of Einstein gravity, with a known mechanism of elementary particles creation, as well.

Non-linearity of Einstein and Lovelock gravity, making it difficult prediction of the properties and possible physical meaning of solutions based on a qualitative analysis of the equations underlying the theory. In this regard, it is important to search the exact solutions and the development of new methods for their generation.

For the case of four-dimensional FRW universe with non-minimal coupling of a scalar field and Gauss-Bonnet scalar, the solutions were obtained in slow-roll approximation~\cite{Guo, Koh,Jiang} and in the case of inflation which is driven by the interaction of a scalar field and GB term without the potential~\cite{Kanti}. But, in the work~\cite{Hikmawan} it was shown that the Gauss-Bonnet inflation without an inflaton potential is not phenomenologically viable. The exact solutions for power-law inflation with nonzero potential were proposed in papers~\cite{Nojiri:2005vv,Guo1,Lahiri,Methew}.

New exact solutions and methods for construction of them are proposed in this article. Special attention we pay to the methods which are based on the confrontation with exact solutions for the model of standard inflation.
The article is organized as follow. In the section 2 we present the model's equation for scalar field cosmology non-minimally coupled to the GB term and showed connection of them with standard FRW inflationary cosmology. In the section 3 we study the possibility of realization of inflationary scenarios of exponent and power law type in the model under consideration. We found such solutions under special choice of scalar field evolution. Section 4 is devoted to search of some method of exact solutions construction with the help of direct connection solutions for the standard inflation with new solutions for the model under consideration imposing the restrictions on a non-minimal interacting term. In the section 5 we demonstrate possible generation of exact solutions from given scale factor in analogy with the standard approach with fine tuning of the potential method. We sum up the results in the section 6.

\section{The model's equations with the Gauss-Bonnet coupling term}
We consider the action with the Gauss-Bonnet term non-minimally coupled to a scalar field~\cite{Guo, Koh}
\begin{eqnarray}\label{S-1}
S = \int_{\mathcal{M}} d^4x\sqrt{-g}\Big[\frac{1}{2} R - \nonumber
 \frac{1}{2}g^{\mu\nu}\partial_{\mu}\phi \partial_{\nu} \phi-\\
 - V(\phi)-\frac12\xi(\phi) R_{\rm GB}^2\Big],
\label{action}
\end{eqnarray}
where $\phi$ is an inflation field with a potential $V(\phi)$, $R$ the Ricci scalar curvature of the spacetime $\mathcal{M}$, $R^{2}_{\rm GB}= R_{\mu\nu\rho\sigma} R^{\mu\nu\rho\sigma} - 4 R_{\mu\nu}R^{\mu\nu} + R^2$  the Gauss-Bonnet  term.
 The Gauss-Bonnet coupling $\xi(\phi)$ is required to be a function of a scalar field in order to give nontrivial effects on the background dynamics.

The background dynamical equations for inflation with the GB term which couples to a scalar field $\phi$ in a spatially flat FRW Universe in the system of units $8\pi G=c=1$ are
\begin{equation}
\label{beq1a}
3H^{2}=\frac{1}{2}\dot{\phi}^{2}+V(\phi)+12\dot{\xi}H^{3}
\end{equation}
\begin{equation}
\label{beq2b}
-2\dot{H}=\dot{\phi}^{2}-4\ddot{\xi}H^{2}-4\dot{\xi}H(2\dot{H}-H^{2})
\end{equation}
\begin{equation}
\label{beq3}
\ddot{\phi} + 3 H \dot{\phi} + V_{,\phi} +12 \xi_{,\phi} H^2 \left(\dot{H}+H^2\right) = 0,
\end{equation}
where a dot represents a derivative with respect to the cosmic time $t$, $H \equiv \dot{a}/a$ denotes the Hubble parameter, and $V_{,\phi} = \partial V/\partial \phi, \,\, \xi_{,\phi} = \partial \xi/\partial \phi$. Since $\xi$ is a function on $\phi$, $\dot{\xi}$ implies $\dot{\xi} = \xi_{,\phi} \dot{\phi}$.

The equation (\ref{beq3}) can be derived from
the equations (\ref{beq1a})--(\ref{beq2b}). Therefore, we will consider the Einstein's -- Friedmann's equations in the following form
\begin{equation}
\label{beq1}
3H^{2}=\frac{\epsilon}{2}\dot{\phi}^{2}+V(\phi)+12\dot{\xi}H^{3}
\end{equation}
\begin{equation}
\label{beq2}
-2\dot{H}=\epsilon\dot{\phi}^{2}-4\ddot{\xi}H^{2}-4\dot{\xi}H(2\dot{H}-H^{2})
\end{equation}
The parameter $\epsilon=1$ for a canonical scalar field and $\epsilon=-1$ for a phantom one.

If $\xi$ is a constant, then equations (\ref{beq1})--(\ref{beq2}) are reduced to those for standard inflationary background equations
\begin{equation}
\label{beq4}
3H^{2}=\frac{\epsilon}{2}\dot{\phi}^{2}+V(\phi)
\end{equation}
\begin{equation}
\label{beq5}
-2\dot{H}=\epsilon\dot{\phi}^{2}
\end{equation}

\section{Direct solutions for inflationary expansions}

Let us start from direct investigation of the equations (\ref{beq1})-(\ref{beq2}).
 Since the equations (\ref{beq1})-(\ref{beq2}) contain four unknown functions, to generate the exact solutions without additional conditions it is necessary to set two of them.

The first step is the selection of the Hubble parameter $H=H(t)$ responsible for inflationary expansion. Then we may choose the scalar field $\phi=\phi(t)$ as the function on time. Substituting this dependence in equation (\ref{beq2}) we find the equation for determination of $\xi$ as the function on $t$. Reverse the relation for $t$ as the function on $\phi$ which we suggested, gives for us possibility to make reconstruction to obtain the dependence $\xi$ on $\phi$. Thus we will have the exact solution of equations (\ref{beq1})-(\ref{beq2}).

\subsection{De Sitter expansion}\label{H=A}
As an example of inflationary expansion we choose the scale factor in the form
\begin{equation}\label{a-exp}
a(t) = a_s \exp\left( At \right),~~A=const.,~~A>0
\end{equation}
where $a_s$ is the constant which means the starting value of the scale factor. Such evolution of the Universe has been applied from the very beginning of inflationary scenarios investigation  \cite{Starobinsky} and still under consideration because of good connection to calculation of cosmological parameters \cite{Planck}.

We can easily find the Hubble parameter
\begin{equation}
H(t)=A\label{H-4-1}
\end{equation}
and set the evolution of the scalar field as
\begin{equation}
\phi(t)=B\exp(-At)\label{phi_4-1},
\end{equation}
where $B=\phi (0)=\phi_{s}$. Let us note that the evolution of the scalar field (\ref{phi_4-1}) widely applied for the chaotic inflation \cite{Linde,Lidsey1991}.

We can proceed with obtaining a solution by the following way.
 By making the substitution (\ref{H-4-1}) 
 into basic equations (\ref{beq1})-(\ref{beq2}) we can find kinetic and potential energy in terms of GB coupling $\xi(\phi(t))$
 \begin{eqnarray}\label{A-phi-xi}
&&\frac{1}{2}\dot{\phi}^2=2A^2\left(\ddot{\xi}-A \dot{\xi}\right) \\
\label{A-V-xi}
&&V(\phi)=A^2\left( 3-2\ddot{\xi}-10 A \dot{\xi}\right)
\end{eqnarray}

Then we can substitute the solution for $\phi$ (\ref{phi_4-1}) into (\ref{A-phi-xi}) to obtain the equation on the function $\xi (t)$:
\begin{equation}
\ddot{\xi}-A\dot{\xi}-\frac{\epsilon B^2}{4}\exp (-2At)=0
\end{equation}

The solution of this equation can be presented in terms of cosmic time $t$ as
\begin{equation}\label{A-xi-t}
\xi=\frac{\epsilon B^2}{24A^2}\exp (-2At)+\frac{C_1}{A}\exp (At) + const.
\end{equation}
We can set constant equal to zero as it gives income into Einstein's part of the gravity equations. Making reconstruction of $\phi$ from
$ \exp \left(At\right)=B/\phi$ we obtain
\begin{equation}
\xi=\frac{\epsilon}{24A^2}\phi^2+\frac{C_1}{A}\frac{B}{\phi}
\end{equation}
%
Using the solution for GB term (\ref{A-xi-t})
we can find the potential from (\ref{A-V-xi})
\begin{equation}
V(\phi)=\frac{\epsilon A^2}{2}\phi^2-12A^3C_1\frac{B}{\phi}+3A^2
\end{equation}
By setting $C_1=0$ we obtain the particular solution and find the analog to the exact solution for a massive scalar field for FRW Universe \cite{givanov1981}, Hubble parameter and the scalar field are still defined by the formulas (\ref{H-4-1}) and (\ref{phi_4-1}), but for standard inflation, the dependences of Hubble parameter and the scalar field from cosmic time differ from
(\ref{H-4-1})-(\ref{phi_4-1})\cite{givanov1981}.
Namely, for the potential of massive scalar fields $ V(\phi)=\frac{m^2 \phi^2}{2}-\frac{m^2}{3\kappa}$ (where $\kappa$ is Einstein's gravitational constant) the scalar field evolution and Hubble parameter where found out in the form
$$
\phi (t)= -m\sqrt{\frac{2}{3\kappa}}t+\phi_0,~~ H(t)=m\sqrt{\frac{\kappa}{6}}\phi
$$
As we can see the advantage of new solution for EGB gravity when $C_1=0$ is enclosed in the case of positive constant term in the potential (cosmological constant) and more reasonable evolution of the scalar field.

\subsection{Power law solution}\label{sect-3-2}
Power law solution is very important for inflationary cosmology as the pattern of exact solution which can solve the horizon, flatness and perturbation-spectrum problems \cite{Lucchin:1984yf}.

To find the exact solution we set
\begin{equation}
H=\frac{m}{t},~~\phi(t)=\sqrt{\frac{2m(c_{1}m+c_{1}+1)}{\epsilon}}\ln t,
\end{equation}
where $m>1$.

Making the procedure described in section \ref{H=A} we obtain the following solution
\begin{equation}
\dot{\xi}=-\frac{c_{1}t}{2m}-c_{2}t^{m+2},\,\, {\xi}(t)=-\frac{c_{1}t^{2}}{4m}-\frac{c_{2}t^{m+3}}{m+3}
\end{equation}
\begin{eqnarray}
\nonumber
V(t)=(5c_{1}m^{2}-c_{1}m+3m^{2}-m)t^{-2}+\\
+12m^{3}c_{2}t^{m-1}
\end{eqnarray}
\begin{eqnarray}
{\xi}(\phi)=-\frac{c_{1}}{4m}e^{2B\phi}-\frac{c_{2}}{m+3}e^{(m+3)B\phi}
\end{eqnarray}
\begin{eqnarray}
\nonumber
V(\phi)=(5c_{1}m^{2}-c_{1}m+3m^{2}-m)e^{-2B\phi}+\\
+12m^{3}c_{2}e^{(m-1)B\phi}
\end{eqnarray}
where $B=\epsilon[2\epsilon m(c_{1}m+c_{1}+1)]^{-1/2}$ and $c_{1}$, $c_{2}$ are arbitrary constants.

Firstly, we note that, in this model, for positive potential $V>0$ we have the conditions
$c_{1}>-(3m-1)/(5m-1)$ and $c_{2}\geq0$.

When $c_{1}=0$ and $c_{2}=0$ we obtain $\xi=0$ and have the case of standard power law inflation.

For $c_{1}=-(3m-1)/(5m-1)$ and $c_{2}=0$ we obtain $V=0$, i.e. the case of not phenomenologically
viable model~\cite{Hikmawan}.

When $c_{2}=0$ and $\epsilon=1$ the solutions correspond to obtained in~\cite{Nojiri:2005vv,Guo1,Lahiri}
in the case of the direct choice of the constants $c_{1}$ and $m$.

Also, we have the additional model with $c_{1}=0$, $c_{2}>0$.

\section{The method of exact solutions construction}

To make possible generation of new solutions for inflation in EGB gravity with non-minimal coupling to a scalar field from well-known solutions in FRW cosmology we will define, as a new approach, the connection between Hubble parameter $ \overline{H}$ of standard FRW inflation and Hubble parameter $H$ in the model with GB term (\ref{S-1}).

In the works~\cite{Guo,Koh,Nojiri:2005vv,Guo1,Lahiri,Methew} there were considered different types of non-minimal coupling, for example, $\xi(\phi)=\xi_{0}\phi^{-n}$. In~\cite{Guo,Koh} the functional dependence of $\xi=\xi(\phi)$ have been introduced a priori and in ~\cite{Nojiri:2005vv,Guo1,Lahiri,Methew} $\xi=\xi(\phi)$ was defined from given Hubble parameter $H(t)$ and a scalar field $\phi(t)$ as functions on time with the aim to find exact solutions.

 We suggest to study the GB coupling $\xi$ as an indicator of Hubble parameter "shifting'' from Hubble parameter $\overline{H}$ of standard inflation to Hubble parameter $H$ in the model (\ref{action}) with GB term non-minimally coupled to a scalar field
\begin{equation}
\label{connection}
\dot{\xi}=\frac{H-\overline{H}}{2H^{2}}
\end{equation}
Then, the equations (\ref{beq1})--(\ref{beq2}) can be rewritten as
\begin{equation}
\label{ex1}
\frac{\epsilon}{2}\dot{\phi}^{2}+V(\phi)=-3H^{2}+6\overline{H}H
\end{equation}
\begin{equation}
\label{ex2}
\frac{\epsilon}{2}\dot{\phi}^{2}=-\dot{\overline{H}}+\overline{H}H-H^{2}
\end{equation}
In the case of $\xi=const$ from (\ref{connection}) one obtains $H=\overline{H}$ and the equations (\ref{ex1})--(\ref{ex2})
are reduced to the equations for standard FRW inflation (\ref{beq4})--(\ref{beq5}). This procedure confirm that $\overline{H}$ is the Hubble parameter for standard inflation.

Further, after simple transformation, we can rewrite the equations (\ref{connection})--(\ref{ex2}) in the following form
\begin{equation}
\label{ex3}
V(\phi)=-2H^{2}+5\overline{H}H+\dot{\overline{H}}
\end{equation}
\begin{equation}
\label{ex4}
\frac{\epsilon}{2}\dot{\phi}^{2}=-\dot{\overline{H}}+\overline{H}H-H^{2}
\end{equation}
\begin{equation}
\label{exi}
\dot{\xi}=\frac{H-\overline{H}}{2H^{2}}
\end{equation}
As can be seen from equation (\ref{exi}), when $\dot{\xi}>0$ the coupling of scalar field to GB term accelerates the expansion of the universe compared to the standard inflation and decelerates in the case of $\dot{\xi}<0$.

Therefore, the GB coupling with arbitrary sign (positive or negative)  can increases or decreases the Hubble expansion rate during inflation depending on the particular form of the function $\xi=\xi(t)$.

The possible way of generating of exact solutions can be associated with selecting of the Hubble parameter $H=H(t)$ and the scalar field $\phi=\phi(t)$ and using the equations (\ref{connection}) and (\ref{ex2}) to define the non-minimal coupling to GB term parameter $\xi(\phi)$. After that we able to generate the exact solutions from equations (\ref{ex3})--(\ref{ex4}) using $\overline{H}$ obtained from (\ref{exi}).

From the the other hand we can go by more natural way. Namely, we can set the Hubble parameter $H(t)$ which give the inflationary expansion and set the dependance $\xi $ on $t$. After that the solution will be defined by the formulae (\ref{ex3})-(\ref{ex4}).

To show the method described above in action, we choose few scale factors for which exact solution exists in terms of elementary functions; inflationary solutions are also there.

\subsection{Class of the exponential solutions}

Let us choose the evolution of the scale factor in the exponential form (\ref{a-exp}). Then we have simple relation for the Hubble constant
$$
H(t)=A, ~~A>0,~~A=const.
$$

It is clear that for given evolution of the GB coupling term $\xi=\xi(t)$ we may found the exact solution making substitution $\xi(t)$ into (\ref{connection}) and then substitute $H$ and $\overline{H} $ into (\ref{ex3})-(\ref{ex4}). Therefore the solution founded in section \ref{H=A} by direct analysis of equations (\ref{beq1})-(\ref{beq2}) does not unique; for various GB coupling terms $\xi(t)$ we may found different exact solutions.

Let us continue to analyse the exponential solution with given GB coupling $\xi$.

\subsubsection{$\xi = c_1 t$}
In the case of linear dependence $\xi $ on time $t$:
\begin{equation}\label{xi-1}
\xi = c_1 t
\end{equation}
after performing the procedure described above one can write the solution
\begin{equation}
\overline{H}=A(1-2c_1A)=const.
\end{equation}
\begin{equation}\label{phi-t-4}
\phi(t)=\pm 2\sqrt{c_1}A\sqrt{A}t+\phi_s
\end{equation}
\begin{equation}
V(t)=2A^2\left(3-5c_1A\right)=const.
\end{equation}

To make connection to the model's action (\ref{action}) we have to define the dependence $\xi$ on $\phi$. To this end we can express $t$ as the function on $\phi$ using the relation (\ref{phi-t-4}). Then we will substitute it into (\ref{xi-1}). The result is
\begin{equation}
\xi=\pm \frac{\sqrt{c_1}}{2A\sqrt{A}}(\phi-\phi_s)
\end{equation}

Thus, by choosing the linear dependence $\xi$ on $t$ we showed that for exponential evolution in FRW cosmology there is exponential solution in EGB cosmology as well.

\subsubsection{$\xi = c_2 t^2$}

Using the quadratic dependence $\xi$ on $t$
\begin{equation}
\xi = c_2 t^2
\end{equation}
we easily can find
\begin{equation}
\overline{H}=A(1-4c_2At),~~ \dot{\overline{H}}=-4A^2c_2
\end{equation}
In this case the potential will not be a constant
\begin{equation}
V(t)=A^2\left[ 3-4c_2-20Ac_2t\right]
\end{equation}
Integrating (\ref{ex4}) we can find
\begin{equation}
\phi (t)= \mp \frac{4}{3}\sqrt{2c_2}(1-At)^{3/2}+\phi_s
\end{equation}
Using this relation we can finally find the dependence $V$ on $\phi$
\begin{equation}
V(\phi)=A^2\Big[ 3-4c_2-20c_2\Big(1\pm \left[ \frac{3}{4\sqrt{2c_2}}(\phi-\phi_s\right]\Big)^{2/3}\Big]
\end{equation}

\subsubsection{$\xi = c_3 t^3$}

Let us take into consideration more complicated case when
\begin{equation}
\xi = c_3 t^3
\end{equation}
Then
\begin{equation}
\overline{H}=A(1-6c_3At^2),~~ \dot{\overline{H}}=-12A^2c_3t
\end{equation}
We can find the dependence $V$ on $t$
\begin{equation}
V(t)=3A^2\left(1-10Ac_3t^2-4c_3t \right)
\end{equation}
Unfortunately, the solution for $\phi$ as the function on $t$
\begin{equation}
\nonumber
\phi(t)=\pm 2\sqrt{3c_3}A\left[ \frac{At-1}{2A}\sqrt{2t-A^2t^2}+\right.
\end{equation}
\begin{equation}
\left.(2A^{2/3})^{-1}\arctan\left( \frac{\sqrt{A}(t-1/A)}{\sqrt{2t-At^2}}\right)\right]
\end{equation}
does not give possibility to express the potential $V$ as the explicit function on $\phi$.

In the general case $\xi=ct^{n}$ the scalar field $\phi$ is defined in quadratures, but one can obtain the another exact solutions
with explicit dependence of the scalar field on the time for some other values of $n$ as can be verified by direct calculations.

Thus, we show that the class of solutions for exponential expansion of the universe is very wide and it depends on the choice of GB coupling function $\xi$.

\subsection{The $\dot{\xi} \& \overline{H} $ ansatz}

Another way of generating the exact solutions on the basis of transformation (\ref{connection}) can be realised in the following way. We suggest the connection between $\dot{\xi} $ and $ H$ in (\ref{connection}). Then we calculate $\overline{H}$ from the same equation (\ref{connection}). After that we can find the solution from (\ref{ex3})-(\ref{ex4}) by integrating (\ref{ex4}) and reconstructing the dependence $t$ as the function of $\phi$ to get the dependence $\xi$ and $V$ on $\phi$.

Let us consider the example of such approach.
We can connect $\dot{\xi}$ with known solution $\overline{H}$ for FRW cosmology in the form $\dot{\xi}=-\left(2H\right)^{-1}$. Then we may choose any solution for FRW cosmology and analyse the corresponding result in EGB cosmology with non-minimal GB coupling (\ref{action}). As an example we may study power law inflation
$$
\overline{H}=n/t,~~n>1
$$
Futher, using the relation from (\ref{connection}) $\overline{H}=2H$, we can transform (\ref{ex3})-(\ref{ex4}) to
\begin{eqnarray}
&&V(\phi (t))=2\overline{H}^2+\dot{\overline{H}}\\
&&\epsilon\dot{\phi}^{2}=-2\dot{\overline{H}}+\frac{1}{2}\overline{H}^2
\end{eqnarray}

For the Hubble parameter $\overline{H}=n/t$ to obtain the real solution for the scalar field we must set $\epsilon=1$. After integration we obtain
\begin{eqnarray}
\phi(t)=\pm\sqrt{\frac{n}{2}(n+4)}\ln(t)\\
V(\phi)=n(2n-1)\exp\left(\frac{\mp 2\sqrt{2}\phi}{\sqrt{n(n+4)}}\right)\\
\xi(\phi)=-\frac{1}{2n}\exp\left(\frac{\pm2\sqrt{2}\phi}{\sqrt{n(n+4)}}\right)
\end{eqnarray}
This solution is in agreement with that in the section \ref{sect-3-2} with $n=2m$ and $c_{1}=1$, $c_{2}=0$.

\subsection{The $H \& \overline{H}$ ansatz}
Now, we consider equations (\ref{ex3})--(\ref{ex4}) with $\overline{H}=H+f(t)$. As the result, we have
the system of equations
\begin{eqnarray}
\label{s}
&&V(\phi)=3H^2+5Hf(t)+\dot{H}+\dot{f}(t)\\
\label{s2}
&&\frac{\epsilon}{2}\dot{\phi}^{2}=Hf(t)-\dot{H}-\dot{f}(t)\\
\label{s3}
&&f(t)=-2\dot{\xi}H^{2}
\end{eqnarray}
where the last equation (\ref{s3}) derived from (\ref{connection}).
The function $f(t)$ is defined by setting $\dot{\xi}$ and $H$.

\subsubsection{$\overline{H}=0$}

At first, we consider the case of $f=-H$ for phantom fields $\epsilon=-1$ with equations
\begin{eqnarray}
\label{sp}
&&V(\phi)=-2H^2\\
\label{sp2}
&&\dot{\phi}^{2}=2H^{2}\\
\label{sp3}
&&2\dot{\xi}H=1
\end{eqnarray}
For this case $\overline{H}=0$ and the connection with the standard inflation is absent.

As the example we choose exponential expansion with
\begin{equation}
a(t)=a_{s}\exp(2A\sqrt{t})
\end{equation}
The Hubble parameter is $H=A/\sqrt{t}$ ($A>0$). Performing integration
 in (\ref{sp})-(\ref{sp3}) we obtain
\begin{eqnarray}
&&V(\phi)=-\frac{16A^4}{\phi^{2}}\\
&&\phi(t)=\pm2A\sqrt{2t}\\
&&\xi(\phi)=-\frac{\sqrt{2}\phi^{3}}{96A^{3/2}}
\end{eqnarray}
For the model with
\begin{equation}
a(t)=a_{s}\exp\left(\frac{B}{\sqrt{2}}\sin(At+C)\right)
\end{equation}
and hence
\begin{equation}
H(t)=\frac{AB}{\sqrt{2}}\cos(At+C)\\
\end{equation}
where $A$, $B$ and $C$ are positive constants, the exact solution can be written as
\begin{eqnarray}
&&\phi(t)=B\sin(At+C)\\
&&V(\phi)=A^{2}\phi^{2}-B^{2}A^{2}\\
&&\xi(\phi)=\frac{\sqrt{2}}{4A^{2}B}\ln\left|\frac{B-\phi}{B+\phi}\right|
\end{eqnarray}
Let us note, that this solution has the feature that the inflationary stages alternated with stages of deflation. Nevertheless, for the very early Universe, when $t\rightarrow 0$ we have the exponential expansion $a(t) \propto \exp (ABt/\sqrt{2})$ which characterise the standard inflation.

\subsubsection{$f=\alpha=const.$}

Further, we consider the case of $f=\alpha$, where $\alpha$ is the arbitrary constant and the equations (\ref{s})--(\ref{s3}) can be transformed to
\begin{eqnarray}
&&V(\phi)=3H^2+5\alpha H+\dot{H}\\
&&\frac{\epsilon}{2}\dot{\phi}^{2}=\alpha H-\dot{H}\\
&&\alpha=-2\dot{\xi}H^{2}
\end{eqnarray}
For the canonical scalar field $\epsilon=1$ with
\begin{eqnarray}
\label{HW}
&&H(t)=C\exp(-At)\\
&&\alpha=\frac{A^{2}B^{2}}{8C}-A
\end{eqnarray}
the FRW space corresponding to the this model is
$$
ds^{2}=-dt^2+\exp\left(-\frac{2Ce^{-At}}{A}\right)\left(dr^2+r^2 d\Omega^2\right)
$$
The standard inflationary model with such space was considered in works~\cite{givanov1981,Barrow:1990vx}.

The exact solution can be represented as
\begin{eqnarray}
\label{pW}
\phi(t)=B\exp\left(-\frac{A}{2}t\right)\\
V(\phi)=\frac{\phi^{2}}{4B^{2}}\left(\frac{24C^{2}}{B^{2}}\phi^{2}+5A^{2}B^{2}-48AC\right)\\
\xi(\phi)=\frac{(8C-AB^{2})B^{4}}{32C^{3}}\phi^{-4}
\end{eqnarray}
Thus, we obtain the exact solution in EGB gravity for double-well potential with corresponding Hubble parameter for standard inflation
\begin{equation}
\overline{H}=H+\frac{A^{2}B^{2}}{8C}-A.
\end{equation}
To simplify the resulting formula we take $C=\frac{5}{48}AB^{2}$. Thus, we obtain
\begin{eqnarray}
V(\phi)=\frac{25}{384}A^{2}\phi^{4}\\
\xi(\phi)=-\frac{4.608}{A^{2}}\phi^{-4}
\end{eqnarray}
with Hubble parameter and scalar field are defined by means of (\ref{HW}) and (\ref{pW}).

The another exact solutions for double-well potential with $H(t)=p/t$ and $\phi(t)=Bt^{n}$ were
obtained in the paper~\cite{Methew}.

\section{Generation of EGB solutions from given scale factor}

It is well known method of exact solution construction in FRW cosmology based on evolutionary scale factor definition \cite{Ellis:1990wsa,Chervon:1997yz,Zhuravlev:1998ff}. In EGB cosmology we have one additional parameter -- GB coupling parameter. Therefore we must find a way to include in the method the parameter $\xi$ also.
To this end, we reduce now equation (\ref{s2}) to the equation for standard inflation by imposing the construction
$$
\frac{df(t)}{dt}=Hf(t)
$$
In this case, the system (\ref{s})-- (\ref{s3}) are reduced to
\begin{eqnarray}
\label{st}
&&V(\phi)=3H^2+\dot{H}+6Hf\\
\label{st2}
&&\epsilon\dot{\phi}^{2}=-2\dot{H}\\
\label{st3}
&&\dot{f}=-2\dot{\xi}H^{3}
\end{eqnarray}
It is clear that by definition of $f(t)$ it equals to the scale factor: $f(t)=a(t)$.

To find the exact solution with performing the integration for the scalar field from the equation (\ref{st2}) we choose the scale factor as $a=a_{s}\exp(B\sqrt{t})$. Then we can obtain the solution
\begin{eqnarray}
\phi(t)=\left(64B^{2}t\right)^{1/4}\\
\label{V}
V(\phi)=\frac{8B^{2}}{\phi^{6}}\Big(3a_{s}\phi^{4}\exp \left(\phi^{2}/8\right)+6B^{2}\phi^{2}-16B^{2}\Big)\\ \xi(\phi)=-\frac{a_{s}e^{\phi^{2}/8}}{B^{6}}\left(\frac{B\phi^{6}}{128}-\frac{3}{16}\phi^{4}+3\phi^{2}-24\right)
\end{eqnarray}
Let us note that the potentials similar to the first term in the (\ref{V}) one can meet in supergravity~\cite{Wetterich}.

Now, we can rewrite the system of equations (\ref{st})-- (\ref{st3}) in the form, suitable for exact solutions construction for any type of the scale factor, as
\begin{eqnarray}
\label{Ht}
&&V(t)=\frac{\ddot{a}}{a}+2\frac{\dot{a}^{2}}{a^{2}}+6\dot{a}\\
\label{Ht1}
&&\phi(t)=\pm\sqrt{\frac{2}{\epsilon}}\int\sqrt{-\frac{d^{2}\ln a}{dt^{2}}}dt+\phi_s\\
\label{Ht2}
&&\dot{\xi}=-\frac{a^{3}}{2\dot{a}^{2}}
\end{eqnarray}
Thus, one can generate the exact solutions for the inflation in EGB gravity by the choice of the scale factor $a=a(t)$.
For standard inflation this procedure was proposed in \cite{Ellis:1990wsa} and it have got development in \cite{Chervon:1997yz,Zhuravlev:1998ff}.

\section{Conclusion}

We study EGB cosmology in 4 dimension with a scalar field having non-minimal coupling to GB term. First of all we made sure that inflationary solutions as exponential and power-law ones are allowable, and the scalar field $\phi$, the potential $V(\phi)$ and GB coupling term $\xi (\phi)$ have been found in explicit form using direct analysis of model's equations. Then we concentrated our attention on searching some connection of EGB cosmology with FRW cosmology (\ref{connection}) by suggestion the relation between Hubble parameter in FRW cosmology $\overline{H}$  with the same for EGB cosmology $H$ in the form $\overline{H}=H(1-2\dot{\xi}H)$. As we can see this relation contains the GB coupling $\xi$ and therefore we need to choose two parameters to connect $\overline{H}$ with $H$. Thus, together with the Hubble parameter in EGB cosmology $H$ we may choose the GB coupling parameter $\xi$ as function on time $t$. As the example we demonstrate that such approach leads to a wide class of exact solutions for exponential evolution of the universe. Then, we applied once again the relation (\ref{connection}) for exact solutions construction using special anzatses which give the examples of power-law, generalized exponential solutions and the solution for the double-well potential. Finally, we made connection of one special anzats with evolutionary scale factor and presented the system of equations in the form which allowed us to generate exact solutions (at least in quadratures) of wide class for given scale factor. Thus the analog of fine tuning of the potential method \cite{Ellis:1990wsa,Chervon:1997yz,Zhuravlev:1998ff} we proposed for EGB cosmology with scalar field and non-minimal interaction with GB term.

\section{Acknowledgements}
Extended talk given at the Ulyanovsk International School-Seminar
"Problems of theoretical and observation cosmology -- UISS 2016"\, September 19-30, 2016, Ulyanovsk, Russia.

I.V. Fomin was supported by RFBR grants 16-02-00488 A and 16-08-00618 A.

S.V. Chervon was partly supported by the State order of Ministry of education and science of RF number  2014/391 on the project 1670.

\end{otherlanguage}
\end{document}